\documentstyle[11pt,newpasp,twoside]{article}
\def\edcomment#1{\iffalse\marginpar{\raggedright\sl#1\/}\else\relax\fi}
\reversemarginpar

\begin{document}
\title{Pressure Balance between Thermal and Non-Thermal Plasmas in the 3C129 
Cluster}
 \author{D. E. Harris}
\affil{Smithsonian Astrophysical Observatory, 60 Garden St.,
Cambridge, MA 02138}

\author{H. Krawczynski}
\affil{Physics Dept., Washington University in St. Louis, 1 Brookings
Drive - Campus Box 1105, St. Louis, MO. 63130}

\begin{abstract}

With new Chandra observations of the cluster containing the two radio
galaxies 3C129 and 3C129.1, we have made a fit to the X-ray surface
brightness to obtain thermal pressures.  VLA data at 1.4 GHz have been
obtained to complement previous maps at 0.33 GHz (Lane et al. 2002)  and at
5 and 8 GHz (Taylor et al. 2001).  From these radio data, we are able to
derive the minimum non-thermal pressure of various emitting volumes
along the tail of 3C129 and in the lobes of 3C129.1.

Under the assumption that the non-thermal plasma excludes significant
thermal plasma, we may expect pressure balance for most features since
ram pressure should be important only close to the cores of the
galaxies.  Since we find that the minimum non-thermal pressures are generally
only a factor of a few below estimates of the ambient thermal
pressure, we conclude that it is unlikely that relativistic protons
contribute significantly to the total pressure.  Reasonable
contributions from low energy electrons and filling factors in the
range 0.1 to 1 suffice to achieve pressure balance.
 
Although we do not find strong signatures for the exclusion of hot gas
from the radio structures, we find soft features near the
cores of both galaxies suggestive of cool gas stripping and hard
features associated with radio jets and possibly a leading bow
shock.

\end{abstract}

\section{Goals}

3C129 and 3C129.1 are two radio galaxies in a nearby cluster of
galaxies (fig.~1) which lies close to the galactic plane.  We proposed
Chandra X-ray observations in order to find signatures of hydrodynamic
interactions between the hot intracluster gas and the radio lobes and
tails and to estimate the non-thermal pressures within the radio
structures by determining the external gas pressure and assuming
pressure balance.  It seemed likely that given the long tail of 3C129
($\approx$~0.5~Mpc projected length), that most of the source was
close to the plane of the sky and that projection effects would thus
be minimal.  Since tailed radio galaxies are found only in clusters of
galaxies, it also seemed likely that the projected distance between
the cluster center and 3C129 was not significantly less than the
actual distance: 3C129 was already known to lie close to the edge of
the detected gas distribution (Leahy \& Yin, 2000).  For these
reasons, we believed 3C129 was a good target for detecting the sort of
depression in X-ray surface brightness caused by cavities in the hot
gas coinciding with radio lobes (e.g. Cygnus A, Carilli et al. 1994; and
Hydra A, McNamara et al. 2000).

The reason we wanted to use pressure balance between thermal and non-thermal 
plasmas is that this procedure provides an estimate of the non-thermal 
pressure which can be compared to that obtained from the usual synchrotron 
formulae assuming equipartition between the energies in relativistic 
particles and the magnetic field.  It is then possible to evaluate the 
likelihood for various values of the critical synchrotron parameters.  To do 
this confidently, we require that unmeasurable ram pressures be negligible  
and that by defining various areas of the radio source, we can infer the 
emitting volumes.

We use the redshift of 3C129.1 (z=0.0208, Spinrad, 1975) as the distance 
indicator of the cluster, and with H$_o$=65~km~s$^{-1}$~Mpc$^{-1}$, one 
arcsec corresponds to 450~pc.

\section{Cluster Gas}

The X-ray emission from the cluster gas is shown in fig.~1.  It is
relatively smooth unlike the clumpy distribution seen for the gas
around the radio galaxies Hydra A (McNamara et al. 2000) and M84
(Finoguenov \& Jones, 2001).  We performed a spectral deprojection
analysis (Krawczynski 2002) and $\beta$ model fits to the radial
distributions in two pie sections are shown in fig.~2.  From these
data we can obtain the thermal pressure as a function of distance from
the cluster center which was judged to lie $\approx~1^{\prime}$ SW of
3C129.1.  It seems likely that the cluster has suffered a recent
merger because the gas distribution is significantly elliptical
(figs.~1 and 2); the cluster contains a radio galaxy with a long tail;
and we find no evidence for a cooling flow (Krawczynski, 2002);

\section{Non-thermal Pressures}

The calculation of minimum non-thermal pressure involves 4
components: the assumption of equipartition between the relativistic
particles and the magnetic field; the assumption that the filling
factor is 1; the assumption that protons do not contribute
significantly to the particle energy; and the assumption that most of
the relativistic electrons have been 'counted' when the synchrotron
luminosity is integrated over some frequency range.  If any of these
assumptions are violated, the total non-thermal pressure will be
greater than the minimum value.

While the usual invocation of equipartition leads to the classical 
$\frac{4}{7}$ power for the pressure's dependence on the filling factor and 
the proton energy density, we should not forget that if 
we were to abandon equipartition and argue that we knew the average magnetic 
field strength from some other method (e.g. detection of inverse Compton 
emission from a known photon distribution), then the dependence is
linear.  Compare the basic equation for the total pressure with the
expression for the minimum non-thermal pressure.



\[P~=~nkT + \frac{1}{3}~\frac{c_{12}(1+k)L}{\phi~V~B^{\frac{3}{2}}}~+~\frac{B^2}{8\pi}\]

\smallskip


\noindent
For the field which minimizes the total pressure, B$_{minP}$



\[P_{NT}~=~0.265~[\frac{c_{12}~(1+k)~L}{\phi~V}]^{\frac{4}{7}}\]


\noindent
where\\


$c_{12}~=~1.06~\times~10^{12}~\frac{\nu_1^{(1-2\alpha)/2}~-~\nu_2^{(1-2\alpha)/2}}{\nu_1^{1-\alpha}~-~\nu_2^{1-\alpha}}$

\smallskip

$L~=~4~\pi~D_L^2~(1+z)^{\alpha~-~1}~\int{k_s~\nu^{-\alpha}~d\nu}$

\smallskip

$\phi$ is the filling factor for the emitting volume,V; k is the ratio
of particle energy densities (protons to electrons), and $k_s$ is
the amplitude of the radiation power law: $S_{\nu}=k_s~\nu^{-\alpha}$.

It is also the case that various conditions can be chosen instead of the 
classical equipartition (particle energy density equals magnetic field 
energy density).  We have chosen to use the magnetic field strength which 
minimizes the total pressure, B$_{minP}$.  Other choices which can change 
the value of the field by factors of up to 1.48 include assuming the field is 
smooth or tangled and equalizing pressures instead of minimizing the total 
pressure, or equalizing the energy densities instead of minimizing the total 
energy (see the appendix of Harris et al. 1995, for further details).

We have chosen regions in 3C129.1 and 3C129 on the basis of minimizing 
uncertainties in converting rectangular areas to cylindrical volumes and 
circular areas to spherical volumes (i.e. we chose regions for which the 
assumption that the depth dimension can be found from transverse dimensions 
is most likely valid).  These regions are shown in fig.~3.

Flux densities for these regions were measured at 0.33, 1.4, 5, and 8
GHz.  Spectral indices were determined for single or broken power laws
from these flux densities as well as from spectral index maps produced
by scaled arrays (5 and 8 GHz, Taylor et al. 2001) and from the
similar uv coverage of our 0.33 and 1.4 GHz data..  Details of this
analysis can be found in Krawczynski et al. (2003).

The results are shown in fig.~4 and given in 
table~1 and it is immediately clear that the minimum 
non-thermal pressures are almost always less than the thermal pressures, and 
that the difference for most regions is a factor of order 0.5.

\begin{table}

\caption{Non-thermal pressures\label{tab:results}}

\begin{tabular}{llcll}
\tableline
Number	& Region	& P$_{min}$	&  P$_{\rm ther}/P_{\rm NT}$ & P$_{\rm ther}/P_{\rm NT}$ \\
	&		& 10$^{-12}$cgs &	$\nu_1=10^7$~Hz	&	$\nu_1=10^6$~Hz \\
\tableline
1	& 129.1: N rectangle	&  6.0		& 5.37		& 2.15		\\
2	& 129.1: N inner	& 21.2		& 1.60		& 1.20		\\
3	& 129.1: S inner	& 24.3		& 1.46		& 0.80		\\
4	& 129: inner arm	&  3.70		& 2.08		& 1.66		\\
5	& 129: W eye		&  2.26		& 3.31		& 2.36		\\
6	& 129: E eye		&  2.30		& 3.45		& 2.79		\\
7	& 129: 4am		&  0.48		& 18.0		& 13.5		\\
8	& 129: 8.9am		&  0.74		& 4.45		& 2.16		\\
9	& 129: 10.5am		&  0.59		& 5.25		& 2.08		\\
10	& 129: 12.7am		&  0.92		& 3.27		& 0.76		\\

\tableline
\tableline

\end{tabular}

Notes to table

The thermal pressures at each location use the temperature from the spectral 
deprojection and the density from the deprojection (weight 2) and $\beta$ 
model fit (weight 1).

The non-thermal pressures were calculated by selecting the magnetic field 
strength which minimizes the total pressure, taking the filling factor, 
$\phi$=1, negligible energy density from protons (k=0); and integrating the 
synchrotron luminosity down to $\nu_1=10^7$ (columns 3 \& 4) or
10$^6$~Hz (column 5).

\end{table}

\section{Implications of Pressure Balance}

Given the uncertainties in calculating various pressures, we find it 
remarkable that the thermal and non-thermal pressures are so close to each 
other for most of the regions selected.  Extending the integration of the 
synchrotron spectrum down to 1 MHz (and thereby including electrons with 
Lorentz factors in the range $\gamma$~=~100 to 500) increases the minimum non-thermal 
pressures to values only a factor of two less than the thermal estimates for 
most regions.  This factor of two is easy to obtain for example by invoking 
a value of the filling factor, $\phi~=~\frac{1}{3}$.  Taken at face value, 
the small factors between the thermal and non-thermal pressures indicate 
that we cannot accommodate large values of k (expected if relativistic 
protons are present); significant departures from equipartition (as 
hypothesized in a different context to facilitate explanations of excess euv emission as IC 
emission, Bowyer, this volume); anomalous numbers of low energy electrons 
(i.e. an excess over the extrapolation  to low energies from the  electron 
spectra inferred from the radio data); and/or values of the filling factor 
$<<$ 0.1.

\section{Features and Conclusions}

Although not all cavities will cause a depression in the X-ray surface 
brightness (Clarke et al. 1997), we interpret the absence of evidence for 
cavities associated with the radio structure of 3C129 to mean that the 
filling factor is likely to be something of order  1/3 or 1/10 (e.g. 
filaments of magnetic field and relativistic particles embedded within the 
ambient thermal plasma).  For 3C129.1, the X-ray surface brightness is 
produced by a long integration of emissivities along the line of sight 
through the cluster center and the relatively small volume occupied by the 
radio structures would not be expected to produce an observable change in 
the X-ray surface brightness.

Whilst analyzing the morphology, we discovered X-ray emission coincident 
with the first few arcsec of the northern radio jet of 3C129 (Harris, 
Krawczynski, and Taylor, 2002) and we find hard emission from the region 
next to the nucleus of 3C129.1, which might be associated with a radio jet 
in that source (fig.~5).

We have also constructed hardness ratio maps (H-S)/H+S) with H = 2 to 5 keV 
and S = 1 to 2 keV.  For both radio galaxies, there is a small region of 
significantly softer emission trailing off to the NW from the host galaxies 
(figs.~6 and 7), and in the case of 3C129, there is a harder region to 
the SE of the galaxy core (fig.~8).  We will examine the statistical 
significance of these features in Krawczynski et al. (2003), but the obvious 
interpretation would be a stand-off bow shock heating the gas ahead of 
3C129, and a short trail of cooler ISM being swept out of each galaxy.  
Although there is little debate about the direction of the relative velocity 
between 3C129 and the ambient gas because of the strong bending of both 
radio jets, previous evidence for relative motion for 3C129.1 has been 
marginal at best.  The fact that both soft trails are in the same quadrant, 
is circumstantial evidence in favor of a large scale mass motion of the ICM 
instead of, or in addition to, the classical explanation for the long tail 
of 3C129 as being caused by a large velocity of the galaxy relative to
a stationary cluster gas.

\acknowledgments

Our collaborators W. Lane, N. Kassim, and G. Taylor provided and/or
reduced the radio data used in this analysis and A.G. Willis obtained
new HI data from the DRAO which was used in the spectral deprojection.
The work at the CfA was supported by NASA grant GO1-2135A and contract
NAS8-39073; and at Yale (HK) by NASA grant GO 0-1169X.

References

\noindent
Clarke, D. A., Harris, D. E., and Carilli, C. L. 1997, MNRAS 284, 981
[Cyg A cavities]

\noindent
Carilli, C.L., Perley, R.A., \& Harris, D.E. 1994, MNRAS 270, 173-177.
[Cyg A cavities]

\noindent
Finoguenov, A.  \& Jones, C. 2001 ApJ 547, L107  [M84]

\noindent
Harris, D. E., Willis, A.G., Dewdney, P.E., \& Batty, J. 1995, MNRAS
273, 785-789 [appendix deals with different sorts of equipartition]

\noindent
Krawczynski, H. 2002, ApJ 569, L27 [3C129 - the cluster deprojection]

\noindent
Krawczynski, H., Harris, D. E., Grossman, R., Lane, W., Kassim, N., \& Willis, 
A.G. 2003, MNRAS (submitted) [3C129-pressure; http://xxx.lanl.gov/abs/astro-ph/0302027].

\noindent
Lane,, W. M., Kassim, N. E., Ensslin, T. A., Harris, D. E., and
Perley, R. A.
2002, AJ 123, 2985-2989

\noindent
Leahy,  D. A. \& Yin, D.  2000 MNRAS 313, 617   [ROSAT analysis]

\noindent
McNamara, B. R. et al. 2000, ApJ 534, L135 [Hydra A cavities]

\noindent
Spinrad, H. 1975 ApJ 199, L1

\noindent
Taylor, G. B., Govoni, F., Allen, S. A., \& Fabian, A. C. 2001 \mnras,
326, 2

\begin{figure}
\label{fig:acis}  %
\caption{The X-ray images for a 10ks ACIS-I observation of 3C129.1
(upper panel) and a 30ks observation of 3C129 with ACIS-S..  Both
observations have been divided into two bands: 1-2keV and 2-5keV.
Each band map was divided by the appropriate exposure map and then the
resulting band images were added together.  The data were smoothed
with a Gaussian of FWHM=30 $^{\prime\prime}$ (3C~129.1) and
FWHM=20$^{\prime\prime}$ (3C~129).  For 3C129, the two back
illuminated chips had a constant level subtracted to compensate for
the additional background experienced from these chips.  Contours from
VLA data at 1400 MHz increase by factors of two and commence with
0.001mJy/beam.  The restoring beamsize is
18$^{\prime\prime}~\times~14^{\prime\prime}$ in PA=-14\deg.}
\end{figure}

\begin{figure}
\caption{The modified King model fits to the data in two 90$^{\circ}$ 
pie sections; one to the north and the other to the West.  The core radius 
is $\approx~9^{\prime}$  and the value of $\beta$ is 
$\approx~\frac{2}{3}$.\label{fig:king}}
\end{figure}

\begin{figure}
\caption{Selected regions in 3C129.1 and 3C129 for calculating 
non-thermal pressures.\label{fig:regions}}
\end{figure}

\begin{figure}
\caption{The pressures from the spectral deprojection analysis (filled 
symbols), the beta model fit (solid curve), and the non-thermal pressures 
for $\nu_1$~=~10$^7$~Hz.\label{fig:pres}}
\end{figure}

\begin{figure}
\caption{An X-ray map of 3C129.1 in the 2 to 5 keV band with radio
contours overlayed.  The X-ray data have been divided by the exposure
map and smoothed with a Gaussian of FWHM=4$^{\prime\prime}$.  The
first brightness level visible is at
1.2~$\mu$photons~cm$^{-2}$~s$^{-1}$~pixel$^{-1}$ and the peak brightness
of the feature of interest to the SW of the radio core is at
2.2~$\mu$photons~cm$^{-2}$~s$^{-1}$~pixel$^{-1}$. The radio map is at a
frequency of 4.88 GHz (from Taylor et al. 2001) with contours
increasing by factors of 2.  The first contour level is 0.1 mJy/beam
and the beam is 1.8$^{\prime\prime}$.
\label{fig:1291jet}}
\end{figure}

\begin{figure}
\caption{Hardness ratio map of 3C129.  Darker grey corresponds to
softer emission and the map 
has been smoothed with a Gaussian of FWHM=8$^{\prime\prime}$.  The
lowest greyscale visible has HR~=~-0.017 and the peak value is -0.032.
On a map covering 16 times the area visible here, there are two
features reaching -0.022.  The radio contours are from a 4.7 GHz map
(Taylor et al. 2001).  The lowest contour is 0.4mJy/beam and
successive contour levels increase by factors of 2.  The clean
beamsize is 1.8$^{\prime\prime}$ FWHM.   
\label{fig:129hr}}
\end{figure}

\begin{figure}
\caption{Hardness ratio map of 3C129.1.  The map has been smoothed
with a Gaussian of FWHM=4$^{\prime\prime}$. The feature of interest
(lying just to the NW of the radio core) has a peak value of -0.128.  The
faint greyscale patches have HR values close to -0.04 and the softest
feature on a map which has 16 times the area shown here is -0.06.  The
radio contours are the same as those in
fig.~5.\label{fig:1291hr}}
\end{figure}

\begin{figure}
\caption{A 2 to 5 keV map of 3C129 divided by the exposure map and
smoothed with a Gaussian of FWHM=8$^{\prime\prime}$.  The first
greyscale visible has a brightness of
0.92~$\mu$photons~cm$^{-2}$~s$^{-1}$~pixel$^{-1}$ and the peak value just
to the N of the radio nucleus (the last circular contour) is
2.14~$\mu$photons~cm$^{-2}$~s$^{-1}$~pixel$^{-1}$.  The pixel size is
0.492$^{\prime\prime}$.  Radio contours are the same as in
fig.~6. In addition to emission from the nucleus and the
first segment of the northern jet, there is extended excess emission
preceding the galaxy. \label{fig:129hard}}
\end{figure}

\end{document}